\documentclass[12pt]{article}
\usepackage{oldgerm}
\usepackage{yfonts}
\usepackage{euler}
\usepackage{graphics}
\usepackage{bm}
\usepackage{graphicx}
\usepackage{amsmath}
\usepackage{amssymb}
\usepackage{amstext}
\usepackage{epstopdf}
\usepackage{amscd}
\usepackage{amsfonts}
\usepackage{appendix}
\usepackage{color}
\usepackage{wasysym}
\usepackage{latexsym}
\DeclareMathAlphabet{\EuFrak}{U}{euf}{m}{n}
\DeclareMathAlphabet{\EuScript}{U}{eus}{m}{n}

\newcommand{\nd}{\noindent}

\newcommand{\be}{\begin{equation}}
\newcommand{\ee}{\end{equation}}
\newcommand{\ben}{\begin{eqnarray}}
\newcommand{\een}{\end{eqnarray}}

\title{{\bf Dimensional Regularization of Renyi's Statistical
Mechanics }}
\author{{A. Plastino$^{1,4,5}$, M. C. Rocca$^{1,2,4}$, G. L. Ferri $^3$}, \\
\small{$^1$ Departamento de F\'{\i}sica,
Universidad Nacional de La Plata,}\\
\small{$^2$ Departamento de Matem\'{a}tica,
Universidad Nacional de La Plata,}\\
\small{$^3$Fac. de C. Exactas-National University La Pampa,} \\
\small{Peru y Uruguay, Santa Rosa, La Pampa, Argentina}\\
\small{$^4$ Consejo Nacional de Investigaciones Cient\'{\i}ficas
y Tecnol\'{o}gicas}\\
\small{(IFLP-CCT-CONICET)-C. C. 727, 1900 La Plata -
Argentina}\\\small{$^5$  SThAR - EPFL, Lausanne, Switzerland}}

\date{\today}

\begin{document}

\maketitle

\begin{abstract}

\nd We show that typical Renyi's statistical mechanics' quantifiers exhibit
poles. We are referring to the partition function ${\cal Z}$ and the mean
energy  $<{\cal U}>$.  Renyi's entropy is characterized by a real
parameter $\alpha$. The poles emerge in a numerable set of
rational numbers belonging to the $\alpha-$line. Physical effects
of these poles are studied by appeal to dimensional
regularization, as usual.
Interesting effects are found, as for instance, gravitational
ones.\\
\nd PACS: 05.20.-y, 02.10.-v\\

\end{abstract}

\newpage

\renewcommand{\theequation}{\arabic{section}.\arabic{equation}}

\section{Introduction}

Renyi's  information measure $S_R$ is a generalization of  both
 Hartley's and Shannon's  entropic quantifiers of our ignorance regarding a system's
structural characteristics. $S_R$ is regarded as a quite  important
measure in several science's areas. We may cite, for example,
ecology, quantum information, Heisenberg's XY spin chain model,
theoretical computing, conformal field theory, quantum quenching,
diffusion processes, etc. \cite{0,1,2,3,4,5,6,7,8,9,10,11}.

\nd The Renyi entropy is also relevant in  statistics as signaling
diversity. $S_R$ is defined as \cite{0}:
\begin{equation}
\label{eq2.1} S_R=\frac {1} {1-\alpha}\ln\left( \int\limits_M
P^{\alpha}d\mu\right).
\end{equation}
We will investigate here poles that emerge in computing the most
important  Renyi's statistical quantities for the harmonic oscillator (HO). We wish to
ascertain the physical significance of these poles.

\nd To such an end we appeal to the dimensional regularization methodology of
 Bollini and  Giambiagi
\cite{tq1}, \cite{tq2,tq4,tq5}, plus its generalization, developed in
 \cite{tp1}.  
Dimensional Regularization is one of the most important advances in theoretical physics and is used in several disciplines of it \cite{dr1}-\cite{dr54}.
 
\vskip 3mm
\nd Why the HO? We do {\bf not} mean to unravel HO's peculiarities here. This is a very well known system already, of course, described by $\alpha=1$ (when  $S_R$  becomes Shannon's entropy).   We use the HO because of its simplicity, so that closed-form formulas become available. 
This enormously facilitates our pole-research and we thus obtain indications as of how to proceed in more complex situations. Our present work is an unavoidable preliminary step to be taken before tackling  such situations.  \vskip 3mm
\nd We will separately treat the one, two, and three dimensions cases, as the poles are different for each dimension. We start below with some general considerations and will heavily rely on reference   \cite{epl},  which should be recommended as a useful   prerequisite.

\setcounter{equation}{0}

\section{Theoretical considerations for the $\alpha-$region outside the poles}

\setcounter{equation}{0}

\subsection{$\alpha >  1$}

We showed in  \cite{epl} that the classical Renyi-HO partition
function is, for $\alpha  >1$,
\begin{equation}
\label{ep8.1}
{\cal Z}=\frac {\pi^{\nu}} {\Gamma(\nu)}
\int\limits_0^{\infty} \frac {u^{\nu-1}}
{[1+\beta(1-\alpha)u]_+^{\frac {1} {1-\alpha}}} du.
\end{equation}
The integral therein involved is
\begin{equation}
\label{ep8.2}
{\cal Z}=
\frac {\pi^{\nu}} {[\beta(\alpha-1)]^{\nu}}
\frac {\Gamma\left(\frac {\alpha} {\alpha-1}\right)}
{\Gamma\left(\frac {\alpha} {\alpha-1}+\nu\right)}
\end{equation}
Also, for the mean energy  $<{\cal U}>$, we have from \cite{epl}
\begin{equation}
\label{ep8.3}
<{\cal U}>=\frac {\pi^{\nu}} {\Gamma(\nu){\cal Z}}
\int\limits_0^{\infty} \frac {u^{\nu}}
{[1+\beta(1-\alpha)u]_+^{\frac {1} {1-\alpha}}} du,
\end{equation}
whose result is
\begin{equation}
\label{ep8.4}
<{\cal U}>=\frac {\nu\pi^{\nu}} {{\cal
Z}[\beta(\alpha-1)]^{\nu+1}} \frac {\Gamma\left(\frac {\alpha}
{\alpha-1}\right)} {\Gamma\left(\frac {\alpha} {\alpha-1}+\nu+1\right)},
\end{equation}
or equivalently,
\begin{equation}
\label{ep8.5}
<{\cal U}>=\frac {\nu}
{\beta[\alpha+\nu(\alpha-1)]}
\end{equation}
The entropy can be expressed via ${\cal Z}$ and $<{\cal U}>$ as
\begin{equation}
\label{ep8.6}
{\cal S}=\ln{\cal Z}+\frac {1} {1-\alpha}
\ln[1+(1-\alpha)\beta<{\cal U}>.
\end{equation}
Using  (\ref{ep2.2}) - (\ref{ep2.4}) we cast  ${\cal S}$ as
\begin{equation}
\label{ep8.7}
{\cal S}=\ln\left\{\left[\frac {\pi} {
\beta(\alpha-1)}\right]^{\nu} \frac {\Gamma\left(\frac {\alpha}
{\alpha-1}\right)} {\Gamma\left(\frac {\alpha}
{\alpha-1}+\nu\right)}\right\}+
\frac {1} {1-\alpha}
\ln\left[\frac {\alpha}
{[\alpha+\nu(\alpha-1)]}\right]
\end{equation}
We gather from  (\ref{ep8.2}) that  ${\cal Z}$ is positive and
finite for  $\alpha>1$.

\subsection{$0< \alpha <1$}   

 Instead, for  $0<\alpha<1$ one has
\begin{equation}
\label{ep8.8}
{\cal Z}=\frac {\pi^{\nu}} {\Gamma(\nu)}
\int\limits_0^{\infty} \frac {u^{\nu-1}} {[1+\beta(1-\alpha)u]^{\frac
{1} {1-\alpha}}} du,
\end{equation}
and
\begin{equation}
\label{ep8.9}
{\cal Z}= \frac {\pi^{\nu}}
{[\beta(1-\alpha)]^{\nu}} \frac {\Gamma\left(\frac {1}
{1-\alpha}-\nu\right)} {\Gamma\left(\frac {1} {1-\alpha}\right)}.
\end{equation}

\vskip  3mm \nd Let us pass now to the mean energy. For  $<{\cal U}>$ we have
\begin{equation}
\label{ep8.10}
<{\cal U}>=\frac {\pi^{\nu}} {\Gamma(\nu){\cal Z}}
\int\limits_0^{\infty} \frac {u^{\nu}}
{[1+\beta(1-\alpha)u]^{\frac {1} {1-\alpha}}} du,
\end{equation}
or
\begin{equation}
\label{ep8.11}
<{\cal U}>=\frac {\nu\pi^{\nu}} {{\cal
Z}[\beta(1-\alpha)]^{\nu+1}} \frac {\Gamma\left(\frac {1}
{1-\alpha}-\nu-1\right)} {\Gamma\left(\frac {1} {1-\alpha}\right)},
\end{equation}
that can be recast as
\begin{equation}
\label{ep8.12}
<{\cal U}>=\frac {\nu}
{\beta[\alpha-\nu(1-\alpha)]}.
\end{equation}
Here
\begin{equation}
\label{ep8.13}
{\cal S}=\ln\left\{\left[\frac {\pi} {
\beta(1-\alpha)}\right]^{\nu} \frac {\Gamma\left(\frac {1}
{1-\alpha}-\nu\right)} {\Gamma\left(\frac {1}
{1-\alpha}\right)}\right\}+\frac {1} {1-\alpha}
\ln\left[\frac {\alpha} {[\alpha-\nu(1-\alpha)]}\right]
\end{equation}
Here we find poles in the partition function.

\vskip  3mm \nd  Outside these poles we have
\begin{equation}
\label{ep8.14}
\frac {1} {1-\alpha}-\nu<0\;\;\;\;\;;\;\;\;\;
\Gamma\left(\frac {1} {1-\alpha}-\nu\right)>0.
\end{equation}
As a consequence,  the partition function is both positive and finite. We use
the equality
\begin{equation}
\label{ep8.15}
\Gamma\left(\frac {1} {1-\alpha}-\nu\right)=
-\frac {\pi} {\sin\pi\left(\nu-\frac {1} {1-\alpha}\right)
\Gamma\left(\nu+1-\frac {1} {1-\alpha}\right)}
\end{equation}
and ascertain that
\begin{equation}
\label{ep8.16}
\sin\pi\left(\nu-\frac {1} {1-\alpha}\right)<0.
\end{equation}
Thus,
\begin{equation}
\label{ep8.17}
2p+1<\nu-\frac {1} {1-\alpha}<2p+2\;\;\;\;;\;\;\;\;p=0,1,2,3,......
\end{equation}
This chain of inequalities shows that  $\alpha$ and $\nu$ are
related to each other.

\section{The divergences of the theory}

\setcounter{equation}{0}

Remember beforehand the well known fact that a classical entropy is defined only up to an arbitrary constant. 
From  (\ref{ep8.9}), ${\cal Z}$' poles arise when the Gamma
arguments become \cite{old}
\begin{equation}
\label{ep2.1}
\frac {1} {1-\alpha}-\nu=-p\;\;{\rm for} \;\;p=0,1,2,3,......,
\end{equation}
or, equivalently, for
\begin{equation}
\label{ep2.2}
\alpha=\frac {1} {2},\frac {2} {3},\frac {3}
{4},\frac {4} {5},......, \frac {\nu-2} {\nu-1},\frac {\nu-1}
{\nu}.
\end{equation}
For  $<{\cal U}>$'s poles we have
\begin{equation}
\label{ep2.3}
\frac {1} {1-\alpha}-\nu-1=-p\;\;\;{\rm for}\;p=0,1,2,3,......,
\end{equation}
or, equivalently,
\begin{equation}
\label{ep2.4}
\alpha=\frac {1} {2},\frac {2} {3},\frac {3}
{4},\frac {4} {5},......, \frac {\nu-1} {\nu},\frac {\nu} {\nu+1}.
\end{equation}

\section{The one-dimensional scenario}

\setcounter{equation}{0}

In one dimension ${\cal Z}$ is regular and $<{\cal U}>$ has a
singularity at $\alpha=\frac {1} {2}$. For $\alpha\neq\frac {1} {2}$, ${\cal
Z}$ and $<{\cal U}>$ can be easily evaluated. The result is
straightforward
\begin{equation}
\label{ep3.1}
{\cal Z}= \frac {\pi} {\beta\alpha},
\end{equation}
\begin{equation}
\label{ep3.2}
<{\cal U}>=\frac {1} {\beta(2\alpha-1)}.
\end{equation}
As a consequence, we have for ${\cal S}$
\begin{equation}
\label{ep3.3}
{\cal S}=\ln\left(\frac {\pi} {\beta\alpha}\right)+
\frac {1} {1-\alpha}
\ln\left(\frac {\alpha} {2\alpha-1}\right)
\end{equation}
When $\alpha=\frac {1} {2}$, we have for ${\cal Z}$
\begin{equation}
\label{ep3.4}
{\cal Z}= \frac {2\pi} {\beta},
\end{equation}
a regular value.  Regularization is needed then only for $<{\cal
U}>$.

\subsection{Dealing with the divergences}

In order to proceed with such regularizing procedure, the main
idea is to  write $<{\cal U}>$ as a function of the dimension
$\nu$ in the fashion
\begin{equation}
\label{ep3.5}
<{\cal U}>_\nu=\frac {2^{\nu+1}\nu\pi^{\nu}} {{\cal
Z}{\beta}^{\nu+1}} \Gamma(1-\nu),
\end{equation}
and carefully dissect this expression. We note first that
\begin{equation}
\label{ep3.6} \Gamma\left(1-\nu\right)= -\frac {1}
{\nu-1}+\boldsymbol{C}+ \sum\limits_{k=1}^{\infty}b_k(\nu-1)^k,
\end{equation}
where  $\boldsymbol{C}$ is  Euler's constant. Let
\begin{equation}
\label{ep3.7}
f(\nu)=\frac {2^{\nu+1}\nu\pi^{\nu}}
{{\beta}^{\nu+1}}
\end{equation}
The  Laurent expansion of  $f(\nu)$ in  $\nu=1$ is
\begin{equation}
\label{ep3.8} f(\nu)=\frac {4\pi} {\beta^2}+ \frac {4\pi}
{\beta^2}\left[1+\ln\left(\frac {2\pi}
{\beta}\right)\right](\nu-1)+
\sum\limits_{k=2}^{\infty}c_k(\nu-1)^k.
\end{equation}
Using  (\ref{ep3.6}) - (\ref{ep3.8}) we obtain
\begin{equation}
\label{ep3.9} <{\cal U}>_\nu=\frac {1} {{\cal Z}} \left\{\frac
{4\pi} {\beta^2(1-\nu)}+ \frac {4\pi}
{\beta^2}\left[\boldsymbol{C}-1-\ln\left(\frac {2\pi}
{\beta}\right)\right]+
\sum\limits_{k=2}^{\infty}d_k(\nu-1)^k\right\}.
\end{equation}
Use now the  ${\cal Z}$ value of  (\ref{ep3.4}) and find
\begin{equation}
\label{ep3.10} <{\cal U}>_\nu= \frac {2} {\beta(1-\nu)}+ \frac {2}
{\beta}\left[\boldsymbol{C}-1-\ln\left(\frac {2\pi}
{\beta}\right)\right]+ \sum\limits_{k=1}^{\infty}a_k(\nu-1)^k.
\end{equation}
Dimensional regularization's prescriptions assert that the $<{\cal
U}>$-physical value is given by the $\nu-1$-independent term in
\begin{equation} \label{ep3.11} <{\cal U}>= \frac {2}
{\beta}\left[\boldsymbol{C}-1-\ln\left(\frac {2\pi}
{\beta}\right)\right].
\end{equation}
Using then  (\ref{ep3.4}) -  (\ref{ep3.11}) we find
\begin{equation}
\label{ep3.12} {\cal S}=\ln\left\{\frac {2\pi} {\beta}
\left[\boldsymbol{C}-\ln\left(\frac {2\pi}
{\beta}\right)\right]^2\right\}.
\end{equation}

\section{The two-dimensional case} 

\setcounter{equation}{0}

\nd For two dimensions, ${\cal Z}$ has a singularity at $\alpha=\frac
{1} {2}$ and $<{\cal U}>$ has singularities at $\alpha=\frac {1} {2}$
and $\alpha=\frac {2} {3}$. Save for the case of   these singularities,
we can evaluate their values of the main statistical quantities
without the use of dimensional regularization. Thus, we obtain
\begin{equation}
\label{ep4.1}
{\cal Z}=\frac {\pi^2} {\beta^2\alpha(2\alpha-1)},
\end{equation}
\begin{equation}
\label{ep4.2}
<{\cal U}>=\frac {2} {\beta(3\alpha-2)},
\end{equation}
\begin{equation}
\label{ep4.3}
{\cal S}=\ln\left[\frac {\pi^2}
{\beta^2\alpha(2\alpha-1)}\right]+
\frac {1} {1-\alpha}
\ln\left(\frac {\alpha} {3\alpha-2}\right)
\end{equation}

\subsection{The $\alpha=1/2$ pole}

\noindent For $\alpha=\frac {1} {2}$ we must employ the treatment of
the preceding Section, i.e.,  regularize, both  ${\cal Z}$ and
${\cal U}$. We start with ${\cal Z}$
\begin{equation}
\label{ep4.4}
{\cal Z}=\left(\frac {2\pi}
{\beta}\right)^{\nu}\Gamma(2-\nu),
\end{equation}
We recall that
\begin{equation}
\label{ep4.5}
\Gamma\left(2-\nu\right)= -\frac {1}
{\nu-2}+\boldsymbol{C}+ \sum\limits_{k=1}^{\infty}b_k(\nu-2)^k,
\end{equation}
and define
\begin{equation}
\label{ep4.6}
f(\nu)=\frac {2^{\nu}\pi^{\nu}} {{\beta}^{\nu}}.
\end{equation}
The associated Laurent expansion is
\begin{equation}
\label{ep4.7}
f(\nu)=\frac {4\pi^2} {\beta^2}+ \frac {4\pi^2}
{\beta^2}\ln\left(\frac {2\pi} {\beta}\right)(\nu-2)+
\sum\limits_{k=2}^{\infty}c_k(\nu-2)^k.
\end{equation}
Using  (\ref{ep4.5}) - (\ref{ep4.7}) we find
\begin{equation}
\label{ep4.8}
{\cal Z}_\nu=-\frac {4\pi^2} {\beta^2(\nu-2)}+ \frac
{4\pi^2} {\beta^2}\left[\boldsymbol{C}- \ln\left(\frac {2\pi}
{\beta}\right)\right]+ \sum\limits_{k=1}^{\infty}a_k(\nu-2)^k,
\end{equation}
and the physical value for he partition function becomes
\begin{equation}
\label{ep4.9}
{\cal Z}= \frac {4\pi^2}
{\beta^2}\left[\boldsymbol{C}- \ln\left(\frac {2\pi}
{\beta}\right)\right].
\end{equation}
Since  ${\cal Z}$ must be positive, we find the following upper bound for  $T$
\begin{equation}
\label{eq1}
T<\frac {e^{\boldsymbol{C}}} {2\pi k}
\end{equation}
\noindent
For ${\cal U}$ the situation is similar. From
(\ref{ep2.4}) we have
\begin{equation}
\label{ep4.10}
<{\cal U}>=\frac {\nu} {{\cal Z}\pi} \left(\frac
{2\pi} {\beta}\right)^{\nu+1}\Gamma(1-\nu).
\end{equation}
Define
\begin{equation}
\label{ep4.11}
f(\nu)=\frac {\nu 2^{\nu+1}\pi^{\nu}}
{{\beta}^{\nu+1}(1-\nu)}.
\end{equation}
The pertinent  Laurent expansion is
\begin{equation}
\label{ep4.12}
f(\nu)=-\frac {16\pi^2} {\beta^2}+ \frac {16\pi^2}
{\beta^2}\left[\frac {1} {2}- \ln\left(\frac {2\pi}
{\beta}\right)\right](\nu-2)+
\sum\limits_{k=2}^{\infty}c_k(\nu-2)^k.
\end{equation}
From   (\ref{ep4.5}) -  (\ref{ep4.12}) we find
\begin{equation}
\label{ep4.13}
<{\cal U}>_\nu=\frac {1} {{\cal Z}} \left\{\frac
{16\pi^2} {\beta^3(\nu-2)}+ \frac {16\pi^2}
{\beta^3}\left[\ln\left(\frac {2\pi}
{\beta}\right)-\boldsymbol{C}-\frac {1} {2}\right]+
\sum\limits_{k=1}^{\infty}a_k(\nu-2)^k\right\}.
\end{equation}
Thus,  ${\cal U}$'s physical value is (remember \ref{eq1})
\begin{equation}
\label{ep4.14}
<{\cal U}>= \frac {4} {\beta}\;\frac
{\ln\left(\frac {2\pi} {\beta}\right)-\boldsymbol{C}-\frac {1}
{2}} {\boldsymbol{C}-\ln\left(\frac {2\pi} {\beta}\right)}.
\end{equation}
Using   (\ref{ep4.9}) -  (\ref{ep4.14}) we find
\begin{equation}
\label{ep4.15}
{\cal S}=\ln\left\{\frac {4\pi^2} {\beta^2}
\frac {\left[\ln\left(\frac {2\pi}
{\beta}\right)-\boldsymbol{C}-1\right]^2}
{\boldsymbol{C}-\ln\left(\frac {2\pi} {\beta}\right)}\right\}.
\end{equation}

\subsection{The $\alpha=2/3$ pole}

\nd For $\alpha=\frac {2} {3}$, ${\cal Z}$ is finite and $<{\cal U}>$
has a pole. The procedure for finding their physical values is
similar to that for the case $\alpha=\frac {1} {2}$. For this reason,
we
 merely indicate the results obtained for ${\cal Z}$, $<{\cal U}>$, and
${\cal S}$. One finds
\begin{equation}
\label{ep4.16}
{\cal Z}=\frac {9\pi^2} {2\beta^2},
\end{equation}
\begin{equation}
\label{ep4.17}
<{\cal U}>=\frac {6} {\beta}\left[\boldsymbol{C}-\frac {1} {2}-
\ln\left(\frac
{3\pi} {\beta}\right)\right],
\end{equation}
\begin{equation}
\label{ep4.18}
{\cal S}=\ln\left\{\frac {36\pi^2} {\beta^2}
\left[\boldsymbol{C}-
\ln\left(\frac {2\pi} {\beta}\right)\right]^3\right\}
\end{equation}

\section{The three-dimensional instance}

\setcounter{equation}{0}

\nd In three dimensions, ${\cal Z}$ has poles at $\alpha=\frac {1}
{2}$ and $\alpha=\frac {2} {3}$ while $<{\cal U}>$ exhibits them
at $\alpha=\frac {1} {2}$, $\alpha=\frac {2} {3}$, and
$\alpha=\frac {3} {4}$. Outside the poles one has for ${\cal Z}$,
$<{\cal U}>$, and ${\cal S}$, respectively,
\begin{equation}
\label{ep5.1}
{\cal Z}=\frac {\pi^3} {\beta^3\alpha(2\alpha-1)(3\alpha-2)},
\end{equation}
\begin{equation}
\label{ep5.2}
<{\cal U}>=\frac {3} {\beta(4\alpha-3)}.
\end{equation}
\begin{equation}
\label{ep5.3}
{\cal S}=
\ln\left[\frac {\pi^3} {\beta^3\alpha(2\alpha-1)(3\alpha-2)}\right]+
\frac {1} {1-\alpha}\ln\left(\frac {\alpha} {4\alpha-3}\right)
\end{equation}
In this case $\alpha$ should satisfy the condition $\alpha<\frac {5} {4}$
for the mean energy to be a positive quantity. 

\subsection{The $\alpha=1/2$ pole}

\noindent For $\alpha=\frac {1} {2}$ we have
\begin{equation}
\label{ep5.4} {\cal Z}_\nu=\left(\frac {2\pi} {\beta}\right)^{\nu}
\Gamma(2-\nu).
\end{equation}
The Laurent expansion is tackled as above. One finds
\begin{equation}
\label{ep5.5}
{\cal Z}_\nu=-\frac {8\pi^3} {\beta^3(\nu-3)}+
\frac {8\pi^3} {\beta^3}\left[
\ln\left(\frac {2\pi} {\beta}-1-\boldsymbol{C} \right)\right]+
\sum\limits_{k=1}^{\infty}a_k(\nu-3)^k
\end{equation}
From (\ref{ep5.5}) it is easy to obtain the physical value of
${\cal Z}$ as
\begin{equation}
\label{ep5.6}
{\cal Z}=\frac {8\pi^3} {\beta^3}\left[ \ln\left(\frac
{2\pi} {\beta}\right)-1-\boldsymbol{C}\right]
\end{equation}
Since ${\cal Z}$ is positive, one is led to the bound
\begin{equation}
\label{eq2}
T>\frac {e^{\boldsymbol{C}+1}} {2\pi k}
\end{equation}
In a similar vein, we  have for $<{\cal U}>$
\begin{equation}
\label{ep5.7}
<{\cal U}>=\frac {1} {\beta}\;
\frac {\frac {7} {2}+3\boldsymbol{C}-3\ln\left(\frac {2\pi} {\beta}\right)}
{\ln\left(\frac {2\pi} {\beta}\right)-\boldsymbol{C}-1}
\end{equation}
and from (\ref{ep5.6}) and (\ref{ep5.7})
\begin{equation}
\label{ep5.8}
{\cal S}=\ln\left\{\frac {2\pi^3} {\beta^3}\frac
{\left[2\boldsymbol{C}+3-2\ln\left(\frac {2\pi}
{\beta}\right)\right]^2}
{\ln\left(\frac {2\pi} {\beta}\right)-1-\boldsymbol{C}}
\right\}
\end{equation}

\subsection{The $\alpha=2/3$ and $\alpha=3/4$ poles}

\noindent
For $\alpha=\frac {2} {3}$ and $\alpha=\frac {3} {4}$ we give only
the corresponding results, since the calculations are entirely
similar to those for the case $\alpha=\frac {1} {2}$. Thus, for
$,\alpha=\frac {2} {3}$ we have
\begin{equation}
\label{ep5.9}
{\cal Z}=\frac {27\pi^3} {2\beta^3}\left[
\boldsymbol{C}-
\ln\left(\frac {3\pi}
{\beta}\right)\right]
\end{equation}
Here one requires
\begin{equation}
\label{eq3}
T<\frac {e^{\boldsymbol{C}}} {3\pi k}
\end{equation}

\begin{equation}
\label{ep5.10}
<{\cal U}>=\frac {1} {\beta}\;
\frac {9\ln\left(\frac {3\pi} {\beta}\right)-6-9\boldsymbol{C}}
{\boldsymbol{C}-\ln\left(\frac {3\pi} {\beta}\right)}
\end{equation}
\begin{equation}
\label{ep5.11}
{\cal S}=\ln\left\{\frac {27\pi^3} {2\beta^3}\frac {\left[
\ln\left(\frac {9\pi^2} {\beta^2}\right)-2-2\boldsymbol{C}\right]^3}
{\left[c-\ln\left(\frac {2\pi} {\beta}\right)\right]^2}
\right\}
\end{equation}
\noindent
For $\alpha=\frac {3} {4}$ we have
\begin{equation}
\label{ep5.12}
{\cal Z}=\frac {32\pi^3} {3\beta^3},
\end{equation}
\begin{equation}
\label{ep5.13}
<{\cal U}>=\frac {4} {\beta}\left[
3\boldsymbol{C}-1-3
\ln\left(\frac
{4\pi} {\beta}  \right)\right],
\end{equation}
\begin{equation}
\label{ep5.14}
{\cal S}=\ln\left\{\frac {5\pi^3} {\beta^3}
\left[\boldsymbol{C}-
\ln\left(\frac {4\pi} {\beta}\right)\right]^4\right\}.
\end{equation}

\section{Specific Heats}

\setcounter{equation}{0}

We set $k\equiv k_B$.  For $\nu=1$, in the regular case we have
for the specific heat $C$:
\begin{equation}
\label{ep6.1} {\cal C}=\frac {k} {2\alpha-1},
\end{equation}
\noindent For $\nu=2$ one has
\begin{equation}
\label{ep6.2} {\cal C}=\frac {2k} {3\alpha-2},
\end{equation}
\noindent Finally, for $\nu=3$ one ascertains that
\begin{equation}
\label{ep6.3} {\cal C}=\frac {3k} {4\alpha-3},
\end{equation}

\subsection{Specific heats at the poles}

\noindent  For $\nu=1$; $\alpha=\frac {1} {2}$
\begin{equation}
\label{ep6.4}
{\cal C}=2k(\boldsymbol{C}-2-\ln 2\pi kT).
\end{equation}
\noindent
For $\nu=2$; $\alpha=\frac {1} {2}$
\begin{equation}
\label{ep6.5}
{\cal C}=\frac {2k(2\ln 2\pi kT-1-2\boldsymbol{C})}
{\boldsymbol{C}-\ln2\pi kT}
- \frac
{2k} {(\boldsymbol{C}-\ln 2\pi kT)^2}.
\end{equation}
\noindent For $\nu=2$ and  $\alpha=\frac {2} {3}$ one has
\begin{equation}
\label{ep6.6}
{\cal C}=6k\left(\boldsymbol{C}-\frac {3} {2}-\ln3\pi kT\right)
\end{equation}
\noindent For $\nu=3$; $\alpha=\frac {1} {2}$,
\begin{equation}
\label{ep6.7}
{\cal C}=k\frac {3\boldsymbol{C}+\frac {7} {2}-3\ln 2\pi kT}
{\ln2\pi kT-\boldsymbol{C}-1}
- \frac {k}
{2(\ln 2\pi kT-\boldsymbol{C}-1)^2}.
\end{equation}
\noindent For $\nu=3$ and $\alpha=\frac {2} {3}$ one finds
\begin{equation}
\label{ep6.8}
{\cal C}=k\frac {9\ln3\pi kT-6-9\boldsymbol{C}} {\boldsymbol{C}-\ln 3\pi kT}
- \frac
{6k} {(\boldsymbol{C}-\ln3\pi kT)^2}.
\end{equation}
\noindent Finally, for $\nu=3$ and $\alpha=\frac {3} {4}$ we obtain
\begin{equation}
\label{ep6.9}
{\cal C}=4k\left(3\boldsymbol{C}-4-3\ln 4\pi kT\right). 
\end{equation}
\nd   Figs, 1, 2, and 3 plot the mean energy's pole-specific heats, within their
allowed temperature ranges, for one, two, and three dimensions,
respectively. The most distinguished feature emerges in the cases
in which we deal with $<U>-$poles for which $Z$ is regular. We see
in such a case that negative specific heats arise. Such an
occurrence has been associated to self-gravitational systems
\cite{lb}. In turn, Verlinde has associated this type of
systems to an entropic force \cite{verlinde}. It is natural to
conjecture then that such a force may appear at the energy associated poles.

\nd Notice also that temperature ranges are restricted. There is
an $T-$upper bound, and one may wish to remember, in this respect, the notion of  Hagedorn
temperature \cite{witten}, as an example
a temperature's upper bound. In two and three dimensions
there is also a lower bound, so that the system (at the poles)
would be stable only in a limited $T-$range.

\section{Discussion}

\setcounter{equation}{0}

\nd In this work we have appealed to an elementary regularization
procedure to study the poles in the partition function and the mean
energy that appear, for specific, discrete q-values, in Renyi's
statistics of the harmonic oscillator. We studied the thermodynamic behavior at the poles and
found interesting peculiarities. The analysis was made in one, two,
three, and $3$ dimensions. Amongst the pole-traits we emphasize:

\begin{itemize}

\item The poles appear, both in the partition function and the mean energy, for  
$0 < \alpha <1.$

\item These poles ar an artifact of having $\alpha \ne 1$.

\item We have proved that there is an upper bound to the
temperature at the poles, confirming the findings of Ref.
\cite{PP94}, in the sense that, for $\alpha \ne 1$, the heath bath 
of the canonical ensemble must be finite.

\item In some cases, Renyi's' entropies are positive only for a
restricted temperature-range. Lower $T$ bounds seem to be
a new trait discovered here.

\item Negative specific heats, characteristic trait of
self-gravitating systems, are encountered.

\end{itemize}

\nd
Our physical results derive only from statistics,
not from mechanical effects. This fact reminds us of a similar
occurrence in the case of the entropic force conjectured by
Verlinde \cite{verlinde}. 

 \vskip 3mm \nd Indeed, the poles arise only because $\alpha \ne 1$. They are a property of the entropic quantifier, not of the Hamiltonian. Indeed, only for $\alpha \ne 1$ a Gamma function appears in the partition function. This Gamma function may display poles.

 \vskip 3mm \nd Future research should be concerned with cases where it is already known in advance that $\alpha \ne 1$. For these cases, the traits here discovered may acquire some degree of physical     
''reality''.  \vskip 3mm

\nd The importance of the present communication resides in that fact of having disclosed Renyi's entropy traits that could not have been suspected before.

\newpage
\begin{figure}[h]
\begin{center}
\includegraphics[scale=0.6,angle=0]{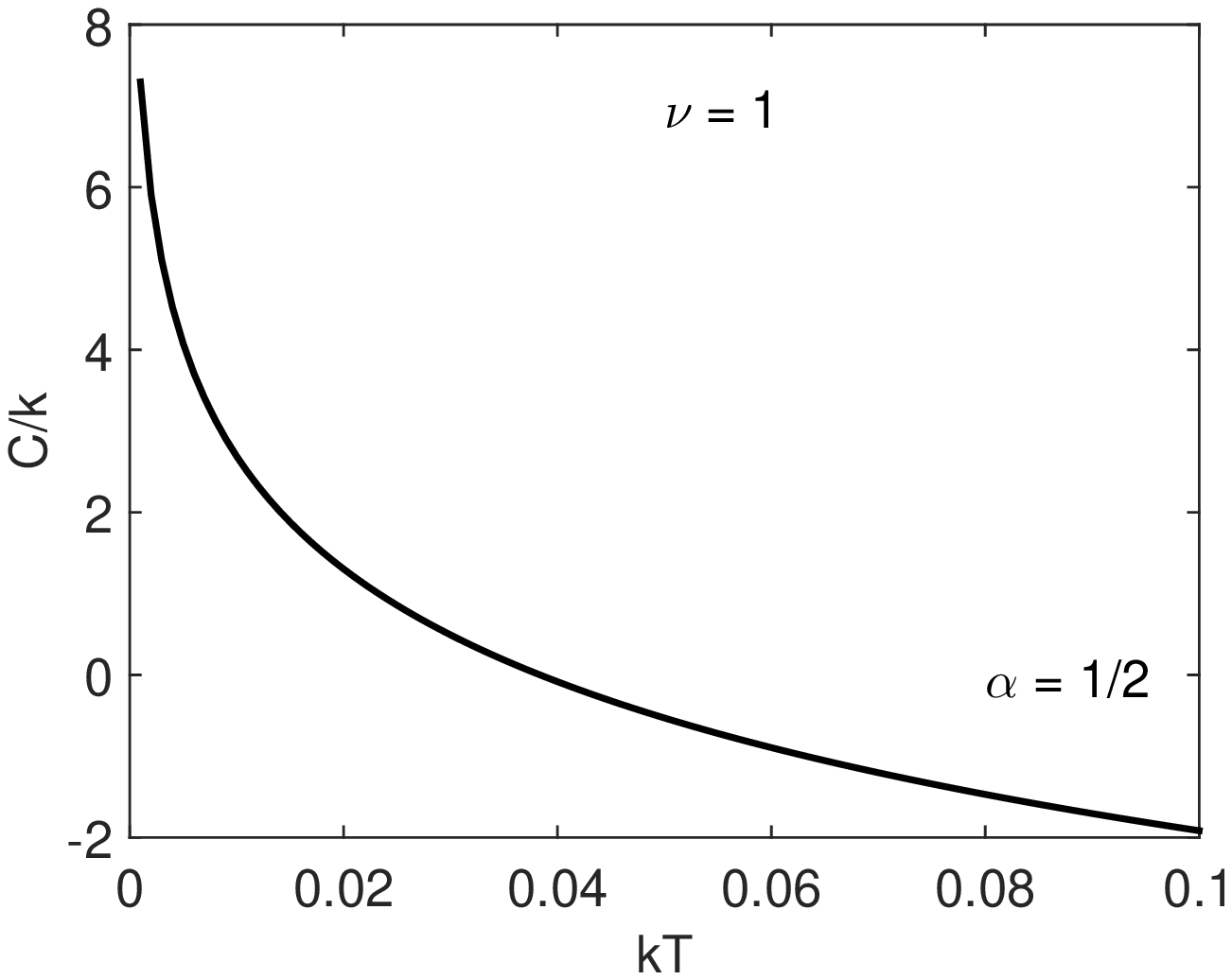}
\vspace{-0.2cm} \caption{One dimension: specific heats at the pole
versus temperature $T$.}\label{fig1}
\end{center}
\end{figure}

\newpage
\begin{figure}[h]
\begin{center}
\includegraphics[scale=0.6,angle=0]{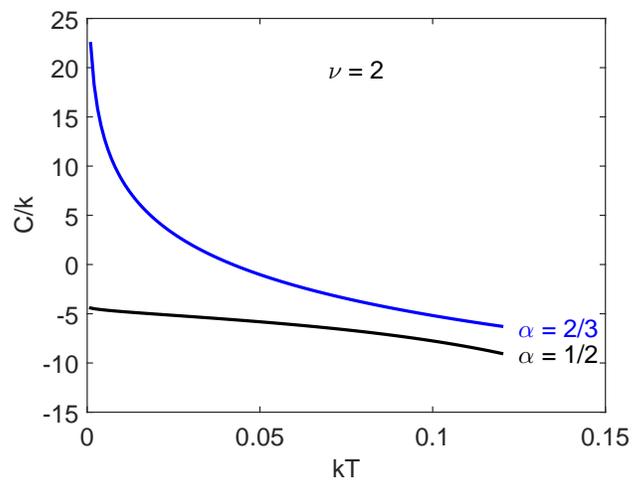}
\vspace{-0.2cm} \caption{Two dimensions: specific heats at the two
poles versus temperature $T$.}\label{fig2}
\end{center}
\end{figure}

\newpage
\begin{figure}[h]
\begin{center}
\includegraphics[scale=0.6,angle=0]{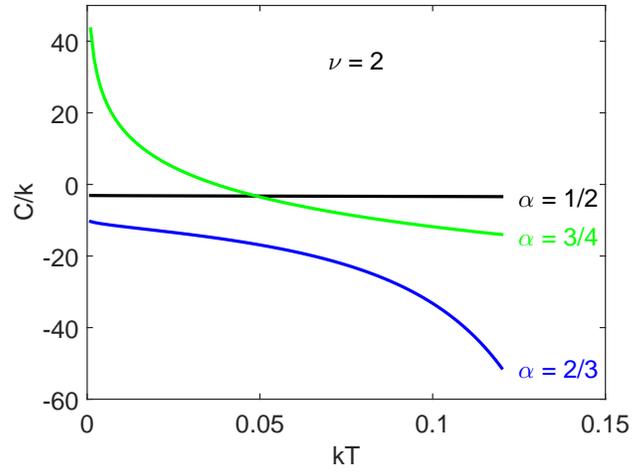}
\vspace{-0.2cm} \caption{Three dimensions: specific heats at the
three poles versus temperature $T$. }\label{fig3}
\end{center}
\end{figure}

\end{document}